\documentclass[pra,aps,amsfonts,amssymb,amsmath,amsmath,twocolumn,showpacs]{revtex4}

\usepackage[utf8]{inputenc}
\usepackage[T2A]{fontenc}
\usepackage[russian,english]{babel}

\usepackage[]{graphicx,color}

\definecolor{darkblue}{rgb}{0,0,0.75}
\definecolor{darkgreen}{rgb}{0.,0.7,0}
\definecolor{darkred}{rgb}{0.7,0,0}
\definecolor{note}{rgb}{0.25,0.0,0.75}
\usepackage[unicode, colorlinks, citecolor=darkblue, linkcolor=darkred, urlcolor=blue]{hyperref}

\begin{document}

\title{Diffraction losses of a Fabry-Perot cavity with nonidentical non-spherical mirrors}

\author{Mikhail V. Poplavskiy}

\affiliation{Faculty of physics, Moscow State University, Moscow 119991, Russia}

\author{Andrey B. Matsko}

\affiliation{Jet Propulsion Laboratory, California Institute of Technology, 4800 Oak Grove Drive, Pasadena, California 91109-8099, USA}

\author{Hiroaki Yamamoto} 

\affiliation{LIGO Laboratory, California Institute of Technology, MC~100-36, Pasadena, CA 91125, USA}

\author{Sergey P. Vyatchanin}

\affiliation{Faculty of physics, Moscow State University, Moscow 119991, Russia,\\
 Quantum Technology Centre, Moscow State University, Moscow 119991, Russia}

\begin{abstract}

Optical cavities with both optimized resonant conditions and high quality factors are important metrological tools. In particular, they are used for laser gravitational wave (GW) detectors.  It is necessary to suppress the parametric instability by damping the resonant conditions of harmful higher order optical modes (HOOM) in order to have high cavity powers in GW detectors. This can be achieved effectively by using non spherical mirrors in symmetric Fabry-Perot (FP) cavities by increasing roundtrip losses of HOOMs  \cite{14FeDeVy,16MaPoYa}. 
Fabry-Perot cavities in most of the GW detectors have non-identical mirrors to optimize clipping losses and reduce thermal noise by reducing the beam size on one side of the cavity facing to the beam splitter and recycling cavities. We here present a general method to design non spherical  non-identical mirrors in non-symmetric FP cavities to damp HOOMs.  The proposed design allows to the suppress the loss of the arm power caused by point absorbers on test masses.

\end{abstract}


\maketitle

\section{Introduction}

In order to study gravitational waves emitted, for instance, due to the merger of binary star systems \cite{GW150914, GW170104, GW170814, GW170817}, demanding sensitivity is required. It calls for improvement of the existing Laser Interferometer Gravitational-Wave Observatory (LIGO), Virgo \cite{virgo}, and the Kamioka Gravitational Wave Detector (KAGRA) \cite{kagra2,kagra1} systems. To achieve the high sensitivity of modern laser gravitational wave (GW) detectors such as Advanced LIGO interferometer (aLIGO) \cite{pfaLIGO, 17LIGO_1, 18LIGO_1} one needs very high circulating optical power. The parametric instability induced by radiation pressure is one of the  causes limiting the power in laser GW detectors of third generation if the mechanical modes of the mirror are not damped.

Optical pumping a Fabry-Perot cavity with mechanical degrees of freedom results in modulation of the pump light at frequencies corresponding to the mechanical modes of the cavity. The optical modulation is coupled to the mechanical motion via the ponderomotive effect and occurs due to a parametric instability \cite{01BrStVy, 02BrStVy}. The parametric instability is caused by interaction of three modes comprising two optical modes of the cavity and one mechanical (acoustic) mode of the cavity mirror when the difference $\omega_0 - \omega_1$ between the frequency $\omega_0$ of the pumped optical mode and the frequency $\omega_1$ of the optical Stokes mode is close to an acoustic mode frequency $\omega_m$ of the cavity mirror. This effect strongly limits the circulating optical power. The opto-mechanic parametric instability phenomenon \cite{01BrStVy} was validated in the table-top experiments involving optical microcavities \cite{15ChZhDa, 05KipVah} as well as in the full-scale gravitational wave detector \cite{01BrStVy, 02BrStVy,EvansPLA2010,15EvGrFr}. 

The system becomes unstable when the pump power circulating in the cavity exceeds a certain threshold value $P_\text{threshold}$  depending on the relaxation rates $\gamma_0$, $\gamma_1$, and $\gamma_m$ of the pumped, Stokes, and acoustic modes, respectively; frequency detuning $\Delta_m =\omega_0-\omega_1-\omega_m$, where $\omega_0$, $\omega_1$, and $\omega_m$ are the frequencies of the pumped, Stokes, and acoustic modes, respectively; geometrical overlap integral $\Lambda$, effective mass of the mechanical mode $M$ and the cavity length $L$.
\begin{equation}
	P_\text{threshold} = \frac{ML^2\omega_m\gamma_m\gamma_1\gamma_0}{\omega_1 \Lambda}\left(1 + \left(\frac{\Delta_m}{\gamma_1}\right)^2\right).
\end{equation}
(This formula is valid for Fabry-Perot cavity, generalization for laser GW detector is presented in \cite{02BrStVy}.)
As the result of the instability the system generates mechanical oscillations at frequency in the vicinity of $\omega_m$ and produces optical harmonics at frequencies $\omega_0 \pm \omega_m$. These harmonics are detrimental in some measurements that involve the cavity so the measurements are usually performed with the optical power not exceeding $P_\text{threshold}$.  

Availability of the multiple high order optical modes as well as multiple mechanical modes in the cavity mirrors increases the probability of the parametric instability. This is especially important for the long base interferometry experiments like LIGO, where condition $|\Delta_m| \sim \gamma_m, \gamma_{0,1}$ can be fulfilled. 

Several techniques increasing the instability threshold in the long cavities were proposed. They include a method of correction of the mirror curvature radius by thermal tuning \cite{07DeZhJu} leading to increase of $|\Delta_m|$. Shifting the higher order optical modes away from the resonance was achieved by heating the non-reflective side of the mirror --- it was successfully applied to mitigate PI was used successfully in $O_1$ run of LIGO. This technique is inefficient with cavities having small coefficient of thermal expansion.

An active optical feedback also suppresses the parametric instability. It can be achieved by injection of a properly prepared light into the Stokes mode \cite{10FaMeZh}. The phase as well as the frequency of the service light should be optimally selected with respect of the pump light. This method cannot be used to suppress a large number of Stokes optical modes. 

Introduced externally electrostatic damping can be utilized to target each mechanical mode reducing its Q-factor \cite{11MiEfBa}. This solution allows suppressing a few elastic modes, it was used successfully in $O_2$ run of LIGO. However, it cannot be used in highly overmoded systems. Despite the fact that the damping scheme does not inject additional thermal noise, this method requires implementation of a separate control loop for each mechanical mode. 

Very recently (before $O_3$ run) similar dampers were designed and applied directly to aLIGO test masses to reduce the quality factors of the unwanted ac modes, while adding a negligible amount of noise.  
The technique calls for attaching several custom dampers to cover the whole frequency range where the unwanted high-Q mechanical mode can appear \cite{LIGO_damper}.

As a more universal method, it was proposed to optimize the cavity mirror shape \cite{14FeDeVy, 16MaPoYa}, leading to an increase in the diffraction loss of all high-order optical modes while keeping low diffraction loss and high Q-factor of the main mode. In such a cavity the parametric instability threshold increases by an order of magnitude. 

The approach \cite{14FeDeVy,16MaPoYa} was developed for a symmetric cavity with identical input and end mirrors. In this paper we generalize it for non-identical mirrors, when axially symmetric but non-spherical input- and end-mirrors differ from each other. This is important for the Advanced LIGO interferometer, because Fabry-Perot cavities in its arms have mirrors with different radius of curvature so that the beam size at the input mirror is slightly less than at the end mirrors. The mirrors are selected in such a way to decrease the diffraction loss at the beam splitter of a limited size. Moreover, by using asymmetric cavity, the beam size can be changed and, by making the beam size on the end mirror larger,one is able to reduce the thermal noise.

The generalization described in this paper is not obvious because the geometrical shape of the optical modes in the optimized cavity is not the Gaussian one and the wave front is not spherical. The system is not self-similar, i.e. wave fronts in different cross sections are non-similar (in contrast with Gaussian beams in which wave fronts are spherical in any cross section). The mirror shape should be optimized to reduce the loss of the main mode while increasing the attenuation of the higher-order modes. The ideal case of the spherical mirror-based cavity was utilized for validation of the numerical technique.

We have developed a numerical technique allowing fast approximation of the optimal shape of the mirror. We used the exact equation then to further optimize the mirror shape. The optimization resulted in increase of the Q-factor of the fundamental mode and reduction of the Q-factor of the other modes.

During the aLIGO and AdVirgo observation runs, many point absorbers were found \cite{PA_aLIGO, PA_AdVirgo}, which affect the arm power and the power recycling.  The causes and the effects were studied and it was shown that the HOOM suppression discussed in this paper does suppress the harmful HOOM related to the performance degradation by the point absorber \cite{PA_HY}. The mirror profile discussed in this paper is not optimal for the point absorber problem, but the same procedure can be used to find the optimal shape for both problems.

This article is organized as follows. In Section II we define a model and describe method of ``propagation of the main mode''. Validity of this technique is verified in Section III for a cavity with spherical mirrors by comparing the results of the numerical simulations with the results of the analytical model. Section IV presents results of the numerical simulations for the case of cavity with non-spherical mirrors. Stability of the numerical results for mirrors characterized with nonzero roughness is studied in Section V. Section VI concludes the paper.

\begin{figure}	
	\includegraphics[width=0.5\textwidth]{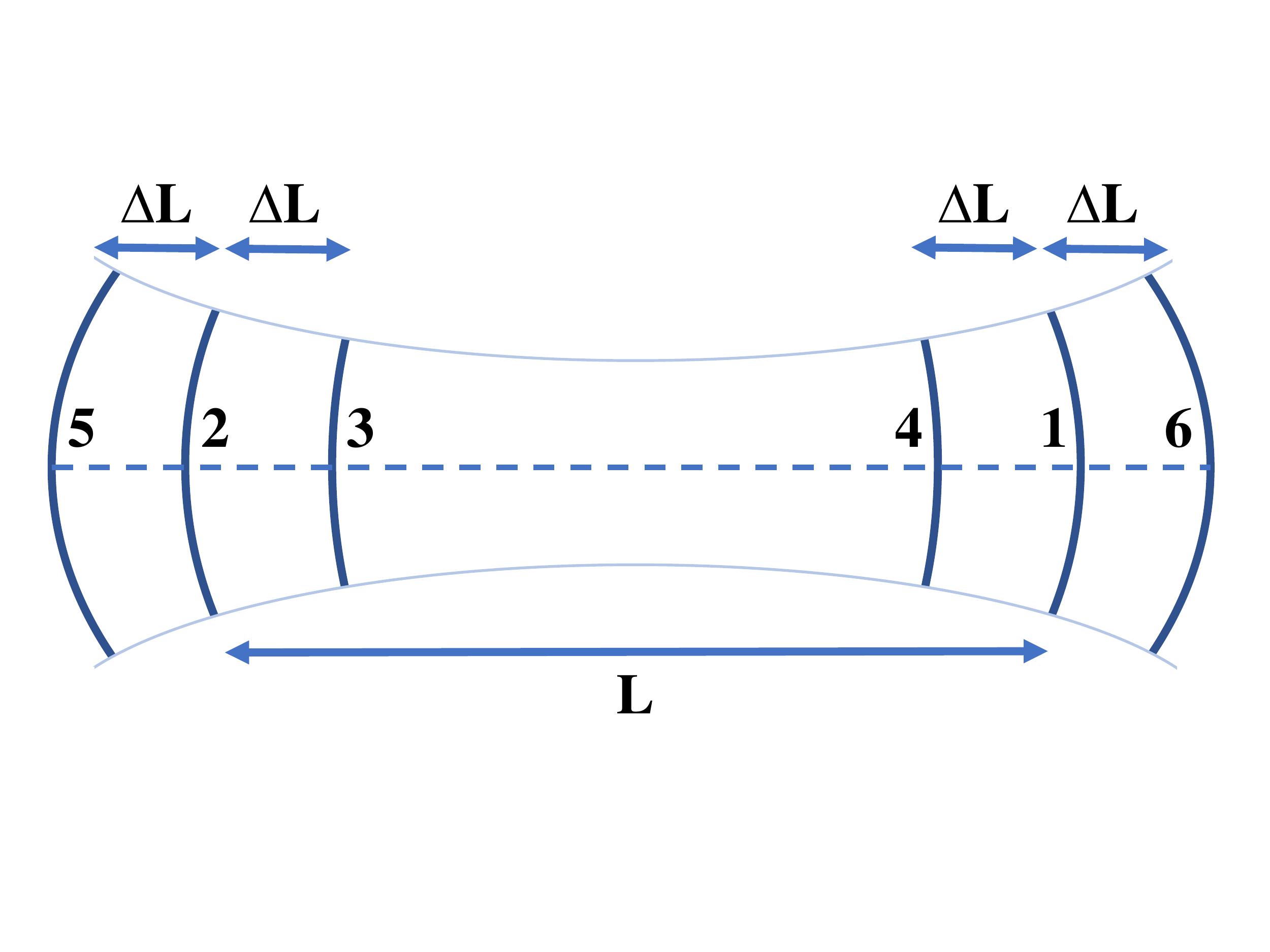}
	\caption{Scheme of a 1D Fabry-Perot cavity. Mirror pairs 5-6, 2-1, and 3-4 represent symmetric cavities. We are interested in finding the optimal mirror shapes for Fabry-Perot cavities with non-identical mirrors, e.g. 3 and 6. While this is a trivial task for the case of spherical mirrors, it becomes computationally intensive when the mirrors are non-spherical. }\label{cav}
\end{figure}

\section{Simulation algorithm}

Modification of the mirror shape in a Fabry-Perot cavity allows reducing the density of the frequency spectrum in the cavity. Our goal is to find the fundamental limitations of the attenuation of a  Fabry-Perot cavity with finite size non-identical mirrors.  The attenuation occurs due to the diffraction loss. While it can be estimated analytically for the case of spherical mirrors, it cannot be found easily for mirrors with arbitrary profile. 

The numerical modeling of a symmetric cavity with identical mirrors involves at least three variables: radius, spatial mirror profile, and mirror reflectivity profile (takes into account a finite mirror size). Consideration of realistic mirrors with distributed roughness also calls for involvement of the angle coordinate into the model. 


In the case of the identical mirrors one needs to consider only one mirror to find the propagator and evaluate the loss due to the finite mirror size. In this case if we use three parameters of profile (25) and curvature radius on centre. Let each of 4 parameters can have 10 values, then we should to look over $10^4$ combinations of parameters (this number is only estimate for example, in reality the number of combination can be smaller if more sophisticated technique is used).
Modeling of a cavity with non-identical mirrors means involvement of two times larger number of variables. Such a computation becomes too long. In example given above we will have 8 parameters, hence, we have to look over $10^8$ (!) combinations of parameters. In addition we can faced with numerical instability if an improper grid is selected. 

We have found that the following three steps allow a significant simplification of the computation problem:
\begin{enumerate}
 \item We  evaluate numerically the main mode field distribution for a Fabry-Perot cavity with identical non-spherical mirrors  \cite{14FeDeVy,16MaPoYa} (mirrors 1 and 2 in Fig.~\ref{cav} with distance $L$ between them). 
 
 \item We evaluate numerically propagation of a wave from the mirror 1 to a short distance $\Delta L$  (much less than the cavity length) to the right {\em outside} of the cavity. Then, calculating the surface of equal phase of wave front, we restore the shape of  the mirror 6 shown in Fig.~\ref{cav}. The size of the beam spot is slightly larger than the initial one. We also calculate propagation of the light from the mirror 2 by the distance $(L+\Delta L)$ to the right. The field distributions on mirror 6 calculated in these two ways should coincide.  It does not due to numerical error. We take a simple average to mitigate the issue.  
 
  \item We evaluate propagation of the wave  to a short distance $\Delta L$ {\em inside} the right from  the mirror 2 and restore its wave front defining the profile of the mirror 3, as shown in  Fig.~\ref{cav}. The size of calculated beam spot is slightly smaller than the initial one. Then we evaluate the wave propagation from the mirror 1 by the distance $(L-\Delta L)$ to the left. The final shape of mirror 3 is found by averaging the shape of the wave fronts obtained in the two calculations.
  
  \item We substitute the approximate solution to the exact equation and further adjust the resultant mirror shape to decrease the attenuation of the fundamental cavity mode and increase the attenuation of the other cavity modes. The step is helpful for reduction of the numerical error. 
\end{enumerate}

As a result of the evaluation we obtain shapes of mirrors 3 and 6 constituting a new Fabry-Perot cavity with non-identical mirrors. We found that the fundamental diffraction loss of this cavity is only slightly different from the diffraction losses of the initial cavity created by the mirrors 1, 2. To validate the calculation we evaluate the phase and amplitude  distributions on mirrors 5 and 4. It should be the same as the distributions for mirror 6 and 3, respectively.

We call the described above procedure as ``the method of mode propagation'' since it utilizes the spatial propagation of the mode of an optimized cavity with identical mirrors to find the optimal shape of the mirrors of an optical cavity with non-identical mirrors. The method allows creating an optical cavity with non-spherical mirrors characterized with low diffraction loss and small laser spot on the input mirror.  The accuracy of this method can be verified using a cavity with spherical mirrors. 

The technique is developed to optimize the computation time. Theoretically, it is possible to find the shape of such a cavity directly, just by fixing the sizes of the beam spots for the front and end mirrors as well as the distance between the mirrors, and by requiring the high finesse for the fundamental mode and low finesse for the rest of the modes. Practically, such an optimization problem involves a large number of variables and cannot be solved utilized existing computer facilities. 

\begin{table}[ht]
\caption{Parameters of a standard Advanced LIGO optical cavity with symmetric spherical mirrors}\label{param}
\begin{tabular}{|l|c|}
\hline
Parameter & Value \\
\hline
Arm length, $L$ & $4$~km\\
Optical wavelength, $\lambda$ & $1064$~nm\\
Intracavity power, $P$ & $800$~kW\\
$AS_{00}$ mode round trip loss, ${\mathcal L}$ & 0.45~ppm\\
$D_{10}$ mode round trip loss, ${\mathcal L}$ & 10~ppm\\
Characteristic cavity length $b=\sqrt{L\lambda/2\pi}$ & $0.0260$~m\\
Radius of mirrors, $R$ &$0.17$~m\\
Dimensionless mirror radius $a_m=R/b$ & $6.53$ \\
Radius $w$ of laser spot at the mirror & $0.06$~m \\
Radius $w_0$ of laser beam at the waist  & $ 0.0115$~m\\
Curvature radius of spherical mirrors, $R_c$ & $2076$~m\\
Geometric parameter $g=1-L/R_c$  of the cavity & $-0.92649$\\
Gouy phase, $\arctan \left[ (b/w_0)^2\right ]$ & 1.378\\
\hline
\end{tabular}
\end{table}

\subsection{Definitions}

A typical Fabry-Perot optical cavity consists of two identical mirrors spaced by distance $L$. We  introduce the following dimensionless coordinates to describe the cavity:
\begin{equation}
  \label{def1}
	x = \frac{r}{b}, \quad b = \sqrt{\frac{L}{k}}, \quad k = \frac{2\pi}{\lambda}, \quad a_m = \frac{R}{b}
\end{equation}   
where $r$ is the distance from the center of the mirror in the plane of the mirror (radial coordinate), $b$ is the scaling factor, $\lambda$ is the optical wavelength, and $R$ is the radius of the mirror. The cavity axis coincide with the mirror axis.

The shape of the cavity mirrors is described by the dimensionless functions
\begin{equation}
h_{1, 2}(x_{1,2}) = k y_{1, 2} (r_{1,2})
\end{equation} 
where $y_{1,2}(r_{1,2})$ is the spatial profile of mirrors (measured in meters) as a function of coordinates $r_{1, 2}$ (measured in meters also). 

A matrix analogue of the Fresnel integral approach gives us a way to evaluate the spatial profile of the cavity eigenmodes and their diffraction loss. This approach is described in detail in \cite{93vinet}, \cite{18PoMaYaVy}. In this paper we present a short summary of the technique.

We define the propagation matrix for the cavity eigenmodes through the Hankel transform:
\begin{equation}
	\textbf{P}^{(\ell)} = \left(\textbf{H}^{(\ell, +)}\right)^{-1} \tilde{\textbf G}(L)\mathbf{H}^{(\ell, +)}
\end{equation}
The matrix representing Green function for the cavity of length $L$ is defined as
\begin{equation}\label{green_func}
\tilde{\textbf G}_{\alpha\beta} (L) = 
  \exp\left(-\frac{i}{2}\cdot \frac{\xi_\alpha^2}{a^2}\right)\delta_{\alpha\beta},\quad
  a =\frac{R_\text{max}}{b},
\end{equation}   
where $R_\text{max}$ is diameter of the circle covered with the numerical grid ($R_\text{max}> R$). The discussion about selection of the numerical grid and is dimensions is presented in what follows. 

The Hankel transform matrix is constructed as
\begin{equation}
\label{Hpm}
	\mathbf H_{\alpha k}^{(\ell, +)} = \frac{2 a^2}{\xi_N^2 \mathcal N_k^{(\ell)}}\, J_{\ell}\left(\frac{\xi_k\xi_\alpha}{\xi_N}\right)
\end{equation}
\begin{equation}	
	 \mathcal{N}_k^{(\ell)} = \left\{ \begin{array}{ll} J_{\ell}^2(\xi_k)\Big(1+\frac{P}{Q\xi_k^2}\Big[\frac{P}{Q}-2\ell\Big]\Big) & \text{if} \quad Q \neq 0, \\
	J_{\ell+1}^2(\xi_k) & \text{if} \quad Q = 0
\end{array} \right.
\end{equation}
where $J_{\ell}$ is the Bessel function of the first kind, $\ell$ is an integer number responsible to order of the mode ($\ell \equiv$ 0 for axial symmetric case), $\xi_n$ is the set of the first $N$ roots of the characteristic equation 
\begin{equation}
\label{eq_root}
P J_{\ell}(x) - Q\, x J_{\ell+1}(x) = 0,
\end{equation}
where $P, \ Q$ are arbitrary numbers.

To account for the diffraction loss, it is assumed that the mode field distribution is limited by a circle of dimensionless radius $a$, which is greater than the radius of the cavity mirror $a_m$. We introduce a window parameter $S$ as
\begin{equation}
  \label{S1}
	S = \frac{a}{a_m}
\end{equation}
The parameter $S$ cannot be selected arbitrarily since the radius of the cavity mirror $a_m$  matches the discrete point belonging to the set $\{\xi_i\}$ of the first $N$ roots of the characteristic equation (\ref{eq_root}). We require the radius of the aperture $a$ to coincide with the boundary point $\xi_N$, and radius of the mirror $a_m$ to coincide with a point $\xi_j$ so that
\begin{equation}
  \label{S}
	S = \frac{\xi_N}{\xi_j}, \quad \quad j < N.
\end{equation}

To fulfill the conditions $a=\xi_N$ and $a_m=\xi_j$ the discrete points of dimensionless x-axis can be selected as
\begin{equation}
	\label{x-axis}
	x_k = \xi_k
\end{equation}

In the case of such a selection our algorithm generates reliable data when the value of the parameter $S$ is in the range of $1.5\le S \le 3$. For the numeric calculations we use $S\simeq 2$, taking into account the condition \eqref{S}. 

The reason for the optimal selection of the parameter $S$ is the increase of the density of the solutions of the characteristic equation with $N$. Increase of $S$ results in the increase of the number of points in the area outside of the mirror where the field is practically absent, that eventually leads to the increase of the numerical error. Our simulation has shown that the number of the points inside the mirror should be twice smaller than the total number of points  $j \sim N/2$  to achieve a reliable result.

The shape of each mirror in the cavity can be introduced by matrices $\mathbf R_{1,2}$ which takes into account the curvature, reflectivity and finite size of the mirrors. For an axial symmetric mirror this matrix is diagonal
\begin{equation}
 (\mathbf R)_{kn} = \exp\big[-ih(x_k)\big]\, D_k\,\delta_{kn}
\end{equation}
where the coefficients $D_{k}$ represent the diaphragm function which sets radius $a_m$ of the mirror
\begin{equation}
\label{diaphragm}
 D_k =\left\{
   \begin{array}{cl}
    1,\quad & \text{if } x_k\le a_m,\\
    0,\quad & \text{if } x_k> a_m
   \end{array}
 \right.
\end{equation}
Finally, we formulated the eigenvalue problem for non-symmetric optical cavity with non-identical mirrors as
\begin{equation}
	\label{EV}
	\left(\mathbf{R_1P\,R_2^2P\,R_1}\right)\Psi = \Lambda\Psi
\end{equation} 
which can be solved numerically. The round trip diffraction loss is $\mathcal{L} = 1 - |\Lambda|^2$, where $\Psi$ is the spatial field distribution on the mirror surface. The azimuth index $\ell$ is not specified for the fundamental modes since $\ell=0$.

Obviously, for a symmetric cavity with the identical mirrors the equation \eqref{EV} is transformed into
\begin{equation}
	\label{EV1}
	\left(\mathbf{R\, P\,R}\right)^2\Psi = \Lambda\Psi
\end{equation} 
where $\mathbf R_1= \mathbf R_2=\mathbf R$, which was used in \cite{16MaPoYa, 18PoMaYaVy}.

\section{Optimization of the numerical algorithm using a cavity with spherical mirrors}

In this section we apply the method of mode propagation to a Fabry-Perot cavity with spherical mirrors. Starting from the cavity with the identical mirrors we calculate the field distribution for a cavity with non-identical ones. The numerical result is validated by the analytic calculations. We have found that the direct propagation of the mode wave front (for example, from  mirror $1$ to mirror $5$ and from mirror $2$ to mirror $6$) gives unacceptable large numerical error as compared with the know theoretical result. Larger mirrors have to be used as an intermediate step in the simulation to overcome the problem, see details in sec. \ref{Sec:ESM}. Alternatively, we can fit the mirror shape and substitute them into the exact numerical model to be able to evaluate and minimize the diffraction loss of the modes.

\subsection{Real sized  mirrors}\label{Sec:MM}

The beam has Gaussian profile in the cavity with infinite spherical mirrors. The same profile is a good approximation if the mirrors are large enough. The deviation occurs in the vicinity of the mirror edge. Let us apply the mode propagation technique to the cavity. We write 
\begin{align}\label{mode_move}
	\tilde{\Psi}_{new} = P(z)\cdot R\cdot \Psi,
\end{align}
where $\tilde{\Psi}_{new}$ is the spatial field distribution on the surface of a mirror placed at coordinate $z$ (phase $\arg\left(\tilde{\Psi}_{new}\right)$ provides information on mirror profile); coordinates of the spherical mirror of an optimal symmetric cavity are $z=0$ and $z=L$; $\Psi$ is the spatial field distribution of eigenmode of the symmetric cavity. We assume that 
\begin{equation}
 \label{z}
 z = {\pm\Delta L;\quad  L \pm \Delta L},
\end{equation}
where the distance change $\Delta L$ is less than the cavity length $L$.

The propagator $P(z)$ can be found from Eq.~(\ref{green_func}) taking into account an arbitrary distance $z$ defined in Eqs.~(\ref{def1}, \ref{z}) due to $a^2\Rightarrow a^2\frac{z}{L}$ \eqref{green_func}. As a result we derive
\begin{equation}
\tilde{\textbf G}_{\alpha\beta} (z) = \exp\left(-\frac{i}{2}\cdot \frac{\xi_\alpha^2}{a^2}\frac{z}{L}\right)\delta_{\alpha\beta}
\end{equation}

 \subsubsection{Symmetric cavity with identical mirrors}
 
Let us consider a symmetric cavity formed by mirrors 1 and 2 (see Fig.~\ref{cav}) with parameters of a standard Advanced LIGO Fabry-Perot optical cavity (see Table~\ref{param}). We need to find the parameters of the the bigger symmetric cavity formed by mirrors 5 and 6 (see Fig.~\ref{cav}). In accordance with Eq.~(\ref{mode_move}) the field distributions are
\begin{align}
	& \tilde{\Psi}_6 = P(L+\Delta L)R \Psi_2 \\
	& \tilde{\Psi}_5 = P(-\Delta L) R \Psi_2
\end{align}
Phase distributions after the mode propagation ($\phi_5^{moved} \equiv $ arg($\tilde{\Psi}_5$) and $\phi_6^{moved} \equiv $ arg($\tilde{\Psi}_6$)) should be the same. The phase distribution defines the mirror shape.

We compare the numerically calculated $\phi_5^{moved},\ \phi_6^{moved}$ (for mirrors 5,6) with analytic ones 
\begin{equation} \label{phianal}
\phi^{analytic} = arg\bigl[\exp \bigl [-i\frac{x^2}{2\rho}\bigr ] \bigr]=-\frac{x^2}{2\rho}	, \quad
\rho = \frac{R^{new}_c}{L}
\end{equation}

The phase distribution is spherical with radius of curvature $R_{c}^{new}$
\begin{align}
	\label{new_rc}
	&R_{c}^{new} = \frac{L_2}{2}\cdot\left(1+\frac{(2\cdot R_c - L)\cdot L}{L_2^2}\right), \\ &\text{where} \quad L_2 = L + 2\cdot \Delta L
\end{align}
The optimal mirror shape follows the phase distribution.

\begin{figure}	
	\includegraphics[width=0.5\textwidth]{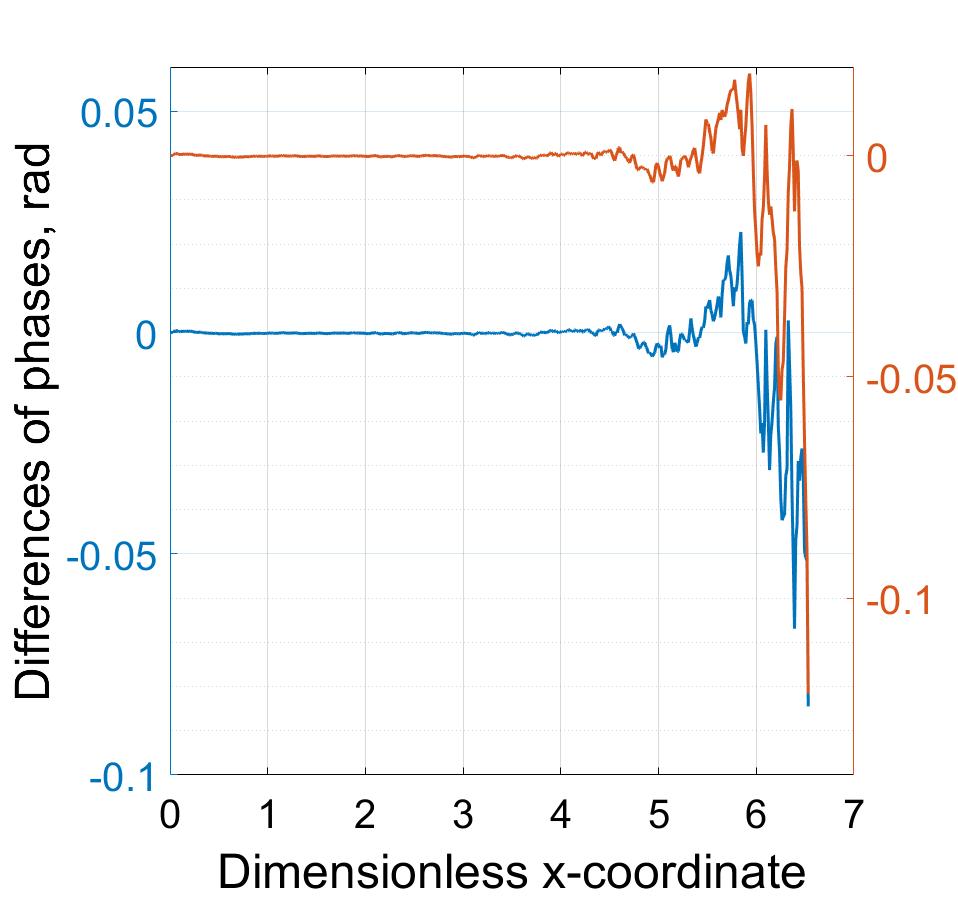}
	\caption{The phase differences (see Fig.~\ref{cav}) found for numerical parameters shown in Table 1 as well as $\Delta L=100$~m. Blue line refers to the phase difference $(\phi_{5}^{analytic} - \phi_{5}^{moved})$ of mirror 5 likewise orange line refers to the same expression for mirror 6 $(\phi_{6}^{analytic} -\phi_{6}^{moved})$ The dimensionless coordinate value $x=6.53$ corresponds to the edge of the mirror with radius 0.17~m.}
	\label{diff_num_th}
\end{figure}

\begin{figure}
\centering
 \includegraphics[width=0.5\textwidth]{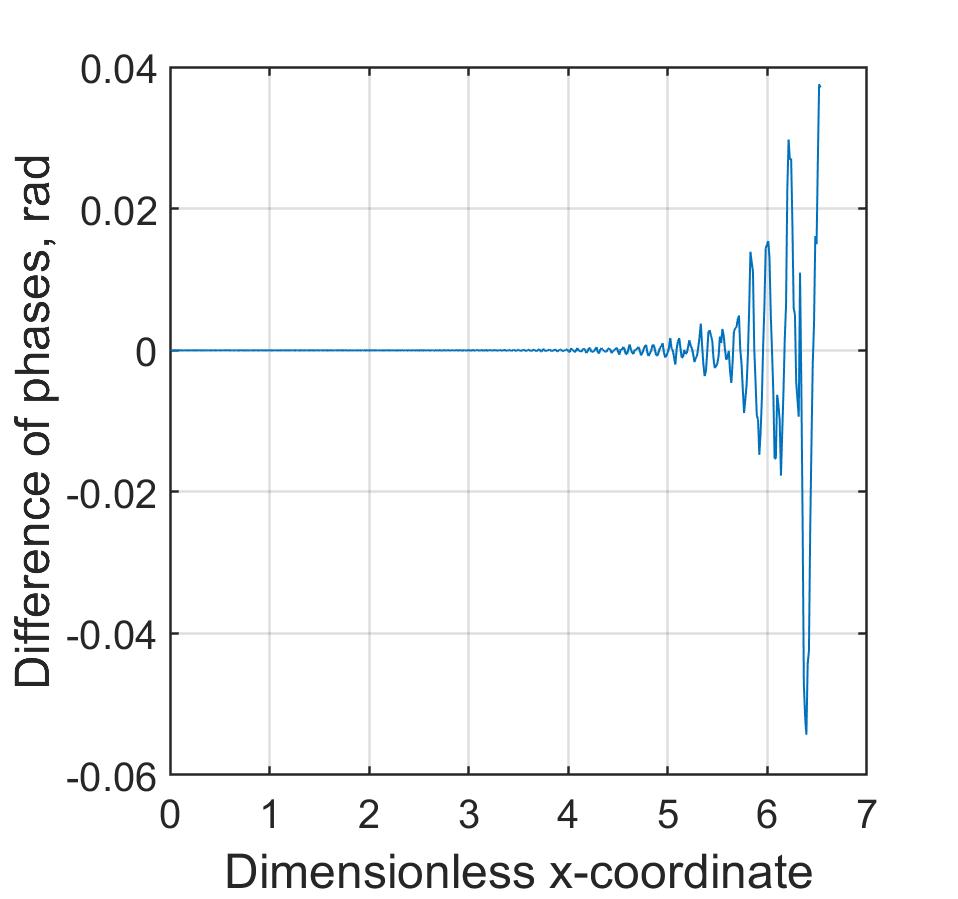}
 \caption{The difference of phases $(\phi_{5}^{moved} -\phi_{6}^{moved})$ of propagated eigenmode (from mirror 1 to mirror 5 and from mirror 1 to mirror 6, see Fig.~\ref{cav}), found for the cavity with mirrors at $x=L+\Delta L = 4100$~m and $x=-\Delta L =- 100$~m, correspondingly using the solution for the cavity with mirrors at $x=L= 4000$~m and $x=0$~m. Here the dimensionless coordinate $x$ is used, and value 6.53 corresponds to the edge of the mirror with radius 0.17 m.}\label{Diff_psi12}
\end{figure}

We find $\phi^{analytic}_5$ and $\phi^{analytic}_6$ from Eq.~(\ref{phianal}) and compare them with the phase distributions of the propagated modes (from mirror 1 to mirror 5 and from mirror 1 to mirror 6 in Fig.~\ref{cav}).  The positions $z$ of the mirrors 5 and 6 are selected to be $z=L+\Delta L = 4100$~m and $z=-\Delta L = -100$~m, correspondingly. The result is shown in Fig.~\ref{diff_num_th}. It is easy to see that the numerical solution deviates from the analytic solution at the edge of the mirror. The numeric solutions found for mirrors 5 and 6 are not identical. The difference between the values of the phase distributions found numerically has a regular component caused by the finite mirror aperture (Fig.~\ref{Diff_psi12}). 

To illustrate the error of the numerical technique, as described above, we consider two cases. In the first case the cavity is composed from mirrors 3 and 6 with analytically evaluated profile.  In the second case the cavity is formed by the mirrors having numerically found profiles (propagation from mirror 2 to mirror 3, located at the distance $\Delta L$, and propagation from mirror 2 to mirror 6, located at distance $L + \Delta L$). 

The results of the calculations are presented in  Table~(\ref{tab_sph_loss}). The evaluated loss for the numerically designed cavity, $\mathcal{L}_1$,  is 3-4 times higher than the loss of the analytically designed cavity, $\mathcal{L}_0$. While this is good accuracy for estimation of the loss of the cavity, we need to improve the result to make it useful for the cavity optimization.

\begin{table}
\centering
\begin{tabular}{|c|c|c|c|c|c|c|c|}
	\hline \\
	Cavity & Length, m & $\Delta L$, m & $\mathcal{L}_0$, ppm &  $\mathcal{L}_1$, ppm \\
	\hline \\
	3-6 & 4000 & 100 & 0.578 & 1.835 \\
	3-6 & 4000 & 150 & 0.724 & 2.665 \\
	\hline
\end{tabular}\caption{Loss for the cavities designed analytically, $\mathcal{L}_0$, and numerically, $\mathcal{L}_1$.}
\label{tab_sph_loss}
\end{table}

While performing the simulations it is important to note that the dimensionless parameters ($x$, $a_m$, $a$ --- in (\ref{def1})) and the matrix elements of the Hankel transform and propagation are normalized by the length $L$ of the initial optical cavity.  The dimensionless coordinate $x$ must be normalized to the new distance $z$ in the case of the mode propagation at a distance different from $L$. The dimensionless coordinate $x$ needs to be re-normalized by the initial distance $L$  upon further simulation of the diffraction losses of the cavity of length $L$ (see Appendix A for details).

We see that even for a symmetric cavity, assembled by mirrors 5 and 6, the method of the mode propagation, as described above, gives unacceptably large numerical errors. To illustrate it we find the attenuation values for a cavity with nonidentical mirrors, 3 and 6, in the next section. We show that the problem can be solved if we first find the numerical solutions for the mirror profile for larger mirrors, and then reduce the mirror size and evaluate the attenuation.

\subsection{Two step evaluation for improved accuracy}\label{Sec:ESM}

The mirror ``cutting'' procedure describe below is aimed to avoid the phase oscillations at the edge of the mirror. These oscillations are due to the fact that the actual eigenmode of the cavity differs from the theoretically found Gaussian mode because the mirror has a finite size.  We increase the mirror radius (for example, from $R=0.17$~m to $R_0=0.3$~m) to obtain the phase distribution after the mode propagation at the some distance $\Delta L$. In this case the oscillations are presented at the edge of enlarged mirror. Then  we cut numerically the excess part of obtained phase distribution and use the profile of the actual smaller mirror to evaluate the attenuation.

The free window parameter $S = a/a_m$ set the ratio between the  modelling dimension $a$ and the mirror size $a_m$. Enlargement of the mirror is determined by changing the window parameter but keeping the area $a$ constant. We select $S' = \xi_N/\xi_M$ ($M < N$) and obtain a new radius of the mirror:
\begin{equation}
	a = a_m \cdot S = a_m' \cdot S', \quad a_m' = a_m \cdot \frac{S}{S'}
\end{equation}

We solve the eigenvalue problem with enlarged spherical mirrors ($a_m' > a_m$), the eigenmodes of the cavity with enlarged mirrors are propagated at the distance $\Delta L$ and cut in size of the initial radius $a_m$. 

\begin{figure}
\centering
 \includegraphics[width=0.5\textwidth]{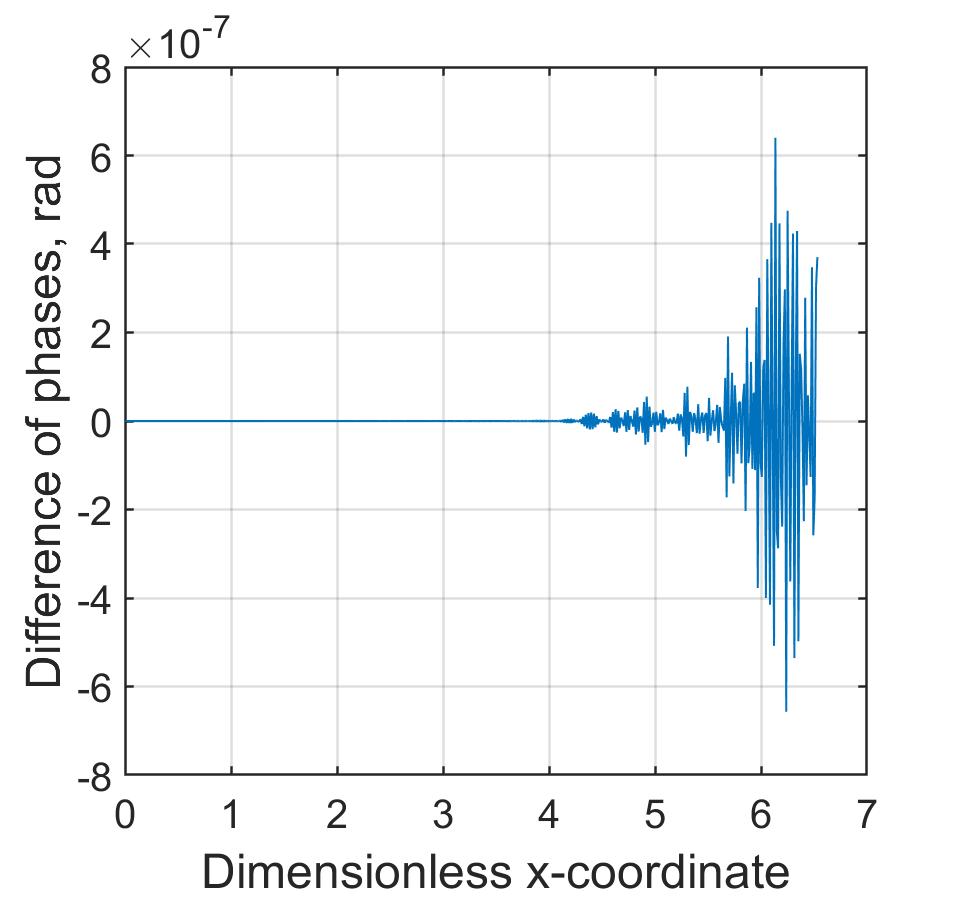}
 \caption{The difference of phases $\phi_{5}^{moved} -\phi_{6}^{moved}$ of the mode propagated at distances $L+\Delta L = 4100$~m and $\Delta L = 100$~m (from mirror 1 to mirror 5 and from mirror 1 to mirror 6, respectively, see Fig.~\ref{cav}), after the "cutting" procedure. Here $x$ value 6.53 corresponds to the edge of the mirror with radius 0.17~m.}\label{Diff_num_psi12}
\end{figure}

For instance, for selection
\begin{align*}
	N = 512,\quad S = 2,\quad S' = 1.3,
\end{align*}
the phase oscillations can be reduced by five orders of magnitude for the original mirror (compare Fig.~\ref{Diff_num_psi12} and Fig.~\ref{Diff_psi12}). The loss evaluation error drops by two orders of magnitude in comparison with the result shown above (compare Table~\ref{tab_sph_loss} and  Table~\ref{Cut_sph}). The selection and subsequent cutting of the larger mirror is not universal. In what follows we see that in the case of the non-spherical mirror the error at the edge of the mirror can result from the very low field there. In that case subsequent numerical optimization gives a better result.

\begin{table}
\centering
\begin{tabular}{|c|c|c|c|c|c|c|c|}
	\hline \\
	Cavity & Length, m & $\Delta L$, m & $\mathcal{L}_0$, ppm &  $\mathcal{L}_2$, ppm \\
	\hline \\
	3-6 & 4000 & 100 & 0.578 & 0.518 \\
	3-6 & 4000 & 150 & 0.724 & 0.726 \\
	\hline
\end{tabular}\caption{The value of the diffraction loss $\mathcal{L}_2$ evaluated for a cavity with nonidentical mirrors using the mode propagation and the "cutting" procedures compared with the numerically evaluated diffraction loss for the analytically found mirror shapes, $\mathcal{L}_0$. }
\label{Cut_sph}
\end{table}

\section{Cavity with non-spherical mirrors}

If utilized in Advanced LIGO, the Fabry-Perot optical cavity with non-spherical mirrors has two advantages over the standard cavity. Firstly, a smaller beam spot at the cavity input mirror (as compared with larger beam spot on end mirror) reduces the diffraction losses at the beam splitter in the interferometer. Secondly, usage of the cavity with non-spherical mirrors mitigates the parametric oscillatory instability by suppressing the high order optical modes (HOOM) by hundreds times \cite{14FeDeVy}, \cite{16MaPoYa}. The beam profile of the fundamental cavity mode is practically Gaussian and, hence, usage of the cavity with the modified mirrors does not require a modification of the auxiliary optics.  In this section we apply the mode propagation technique to the cavity with nonidentical non-spherical mirrors to find both the optimal mirror profile and the associated diffraction loss.

\begin{figure}[h!]
	\includegraphics[width=0.49\textwidth]{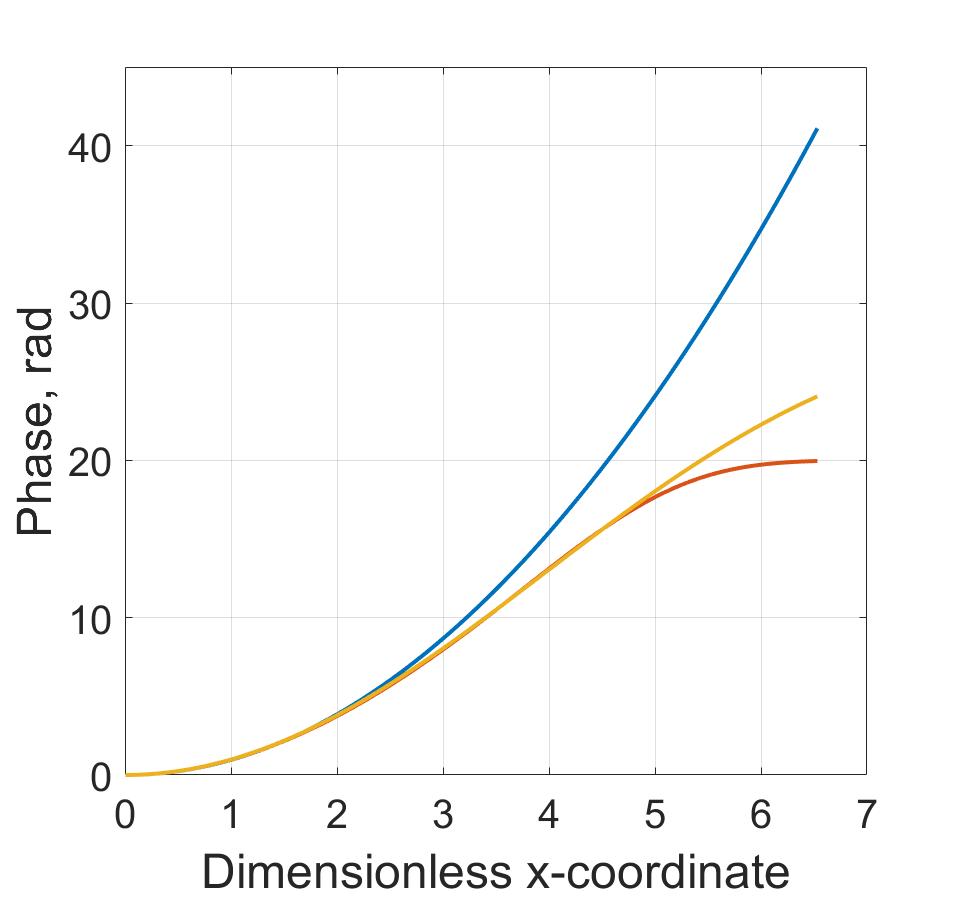}
 \caption{Typical profiles of the mirror surface. Blue line: mirror with spherical profile. Red and yellow lines correspond to parameter sets 1 and 2 (see Table~\ref{EV_problem_asph}).}\label{profiles}
\end{figure}

We consider an optical cavity with the following profile of the mirror surface:
\begin{equation}
	h(x) = h_0\exp^{-\eta\left(1 + \alpha \eta + \beta \eta^2\right)}, \quad \eta = \frac{x^2}{2\rho h_0},
\end{equation}
where $\alpha, \beta, h_0$ are the dimensionless parameters which characterizing the shape of mirrors, $\rho = R_c/L$ is the dimensionless geometric parameter with radius of curvature $R_c$ at the center of mirror (Fig.~\ref{profiles}).

We select several non-spherical shapes and find the diffraction loss for the axial symmetric ($\ell \equiv 0$) eigenmodes (AS00, AS01, AS02) as well as asymmetric dipole ($\ell \equiv 1$) eigenmodes (D10, D11). The result indicates that the HOOM have much higher attenuation than the fundamental mode.
\begin{table}[h!]
\centering
\begin{tabular}{||c||c|c|c||c||c|c|c|c|c|}
 $\#$ & $h_0$ & $\alpha$ & $\beta$ & $w$ cm & $\mathcal{L}_{00}$ & $\mathcal{L}_{01}$ & $\mathcal{L}_{02}$ & $\mathcal{L}_{10}$ & $\mathcal{L}_{11}$ \\
	\hline
	1 & 20 & 0.1525 & 0.35 & 4.857 & 2.198 & 42770 & 46530 & 945 & 20190 \\
	2 & 27.5 & 0.21 & 0 & 4.994 & 2.577 & 18960 & 42270 & 1094 & 40860 \\ 
	3 & 30 & 0.175 & -0.05 & 4.97 & 3.327 & 19610 & 37440 & 1596 & 35380
\end{tabular}\caption{Round trip loss for optical cavities with identical non-spherical mirrors ($R_c, \alpha, \beta, h_0$) in ppm. The radius of curvature for all the parameter sets is equal to 2014~m. The values of the laser spot size ($w$) on the mirror are less than ones in the LIGO cavity (6~cm).}\label{EV_problem_asph}
\end{table}

At the next step we use each parameter set from the Table~\ref{EV_problem_asph} ($h_0, \alpha, \beta$) and smaller cavity length $L - 2\Delta L = 3700$ m for optical cavity with two identical mirrors (corresponds mirrors 3 and 4 on fig.~\ref{cav}). Then we propagate the eigenmode of such cavities to the distances $-2\Delta L = -300$ m and $L = 4000$ m. In this way we have one preliminary specified mirror (for example, mirror 3) by parameters ($h_0, \alpha, \beta$) and another mirror obtained by mode propagation at distances $-2\Delta L$ and $L$.

An averaged phase distribution of the propagated mode in two directions (denote it as $y_{new}(x)$) is irregular at the edges of the mirror because of the numerical error due to the small field amplitude in the area. The mirror cutting technique is inefficient for the non-spherical mirror.  We use a smooth function
\begin{equation}
\label{ynew}
	h_{new}(x) = \tilde{h}_0\left (1-\exp(-\sum_i c_i x^i) \right )
\end{equation}
to optimize the mirror shape to reduce the loss of the fundamental mode by selecting coefficients $c_i$ ($i=1\dots 6$). 

The procedure of optimization is following. 
\begin{enumerate}
\item Parameter $\tilde{h}_0$ is selected manually. 
\item Then expression $\log \left(\frac{\tilde{h}_0}{\tilde{h}_0 - y_{new}(x)}\right)\cdot \frac{1}{x}$ is fitted by polynomial function of 6th degree using standard MatLab tool ``polyfit''. 
\item The obtained coefficients $c_i$ are substituted in Eq. \ref{ynew}. The new mirror profile $h_{new}(x)$ is substituted into the eigenvalue problem (Eq. \ref{EV}) for cavity with non-identical mirrors. 
\item Obtained coefficients $c_i$ and $\tilde{h}_0$ are slightly adjusted for reaching the local minimum of fundamental mode loss for non-symmetrical cavity. Coefficients $c_i$ and $\tilde{h}_0$ obtained after optimization for each of three parameter sets are presented in Table \ref{coef}.
\end{enumerate}

Using \eqref{EV} we evaluated numerically the diffraction loss of the asymmetric cavity eigenmodes. The values of the loss and the effective radii of the main mode at the mirrors are listed in Table \ref{new_loss}.

Both laser spots are smaller ($\simeq 4.7 \div 4.9$ cm and $\simeq 5.4 \div 5.6$ cm) than the laser spot in the standard cavity with spherical mirrors ($\simeq 6$~cm) and the same separation between the mirrors. The diffraction loss of HOOM is much bigger than one of the high order modes of the standard Advanced LIGO cavity. 
\begin{equation}
\label{mismatch}
	I_{mismatch} \sim \int \left(|E_{AS00}| - |E_{Gauss}|\right)^2 x dx
\end{equation}
The integral (\ref{mismatch}) allows to estimate mismatch between the fundamental mode of the new cavity and Gaussian with the same laser spot. The mismatch is on the order of $10^{-3}$ for all the three cavities considered in the simulations.

\begin{figure}[h!]
	\includegraphics[width=0.49\textwidth]{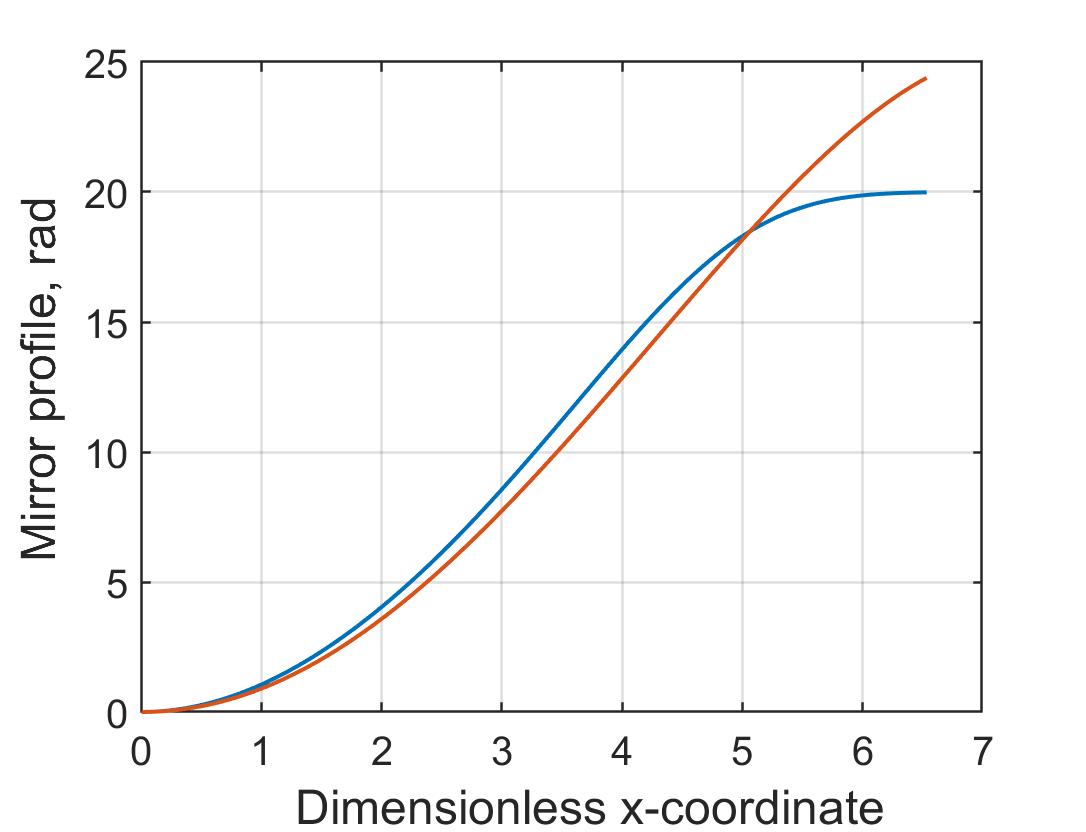}
 \caption{Blue line: phase distribution of preliminary specified mirror 3 (using parameter set 1 from Table~\ref{EV_problem_asph}). Orange line: phase distribution of moved mode (eigenmode of symmetric cavity with two identical non-spherical mirrors corresponding to parameter set 1 and cavity length $L - 2\Delta L = 3700$ m.) at distances $-2\Delta L = -300$ m and $L = 4000$ m. }\label{Mir_as1_new_loss}
\end{figure}

\begin{table}[h!]
\centering
\begin{tabular}{|c|c|c|c|c|c|c|c|}
 $\#$ &$w_1$ & $w_2$ & $\mathcal{L}_{00}$ & $\mathcal{L}_{01}$ & $\mathcal{L}_{02}$ & $\mathcal{L}_{10}$ & $\mathcal{L}_{11}$ \\
	\hline
	1 & 4.699 & 5.407 & 2.765 & 31374 & 25839 & 502.7 & 74572 \\
	2 & 4.858 & 5.567 & 1.323 & 45669 & 55515 & 256.0 & 57819 \\ 
	3 & 4.849 & 5.536 & 0.998 & 50391 & 49940 & 217.6 & 48887 
\end{tabular}\caption{Round trip loss for three different optical cavities with non-spherical mirrors (in ppm) after mode propagation at the distances $-2\Delta L = 300$~m and $L = 4000$~m and after re-optimization of obtained phase distribution.  The radii $w_1$ and $w_2$ of the beam spots at the mirrors are shown for the main mode.}\label{new_loss}
\end{table}

\section{Sensitivity to the small roughness of the mirrors}

\begin{figure*}
	\includegraphics[width=0.325\textwidth]{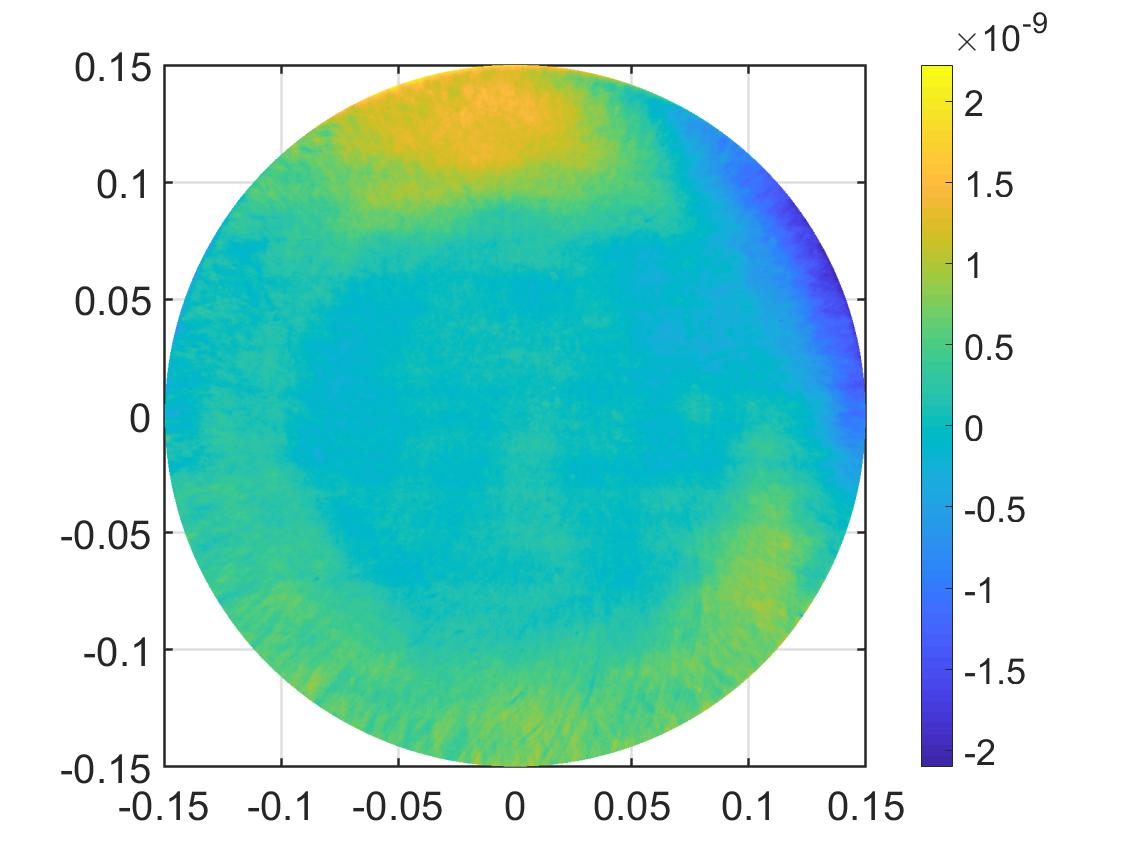}
	\includegraphics[width=0.325\textwidth]{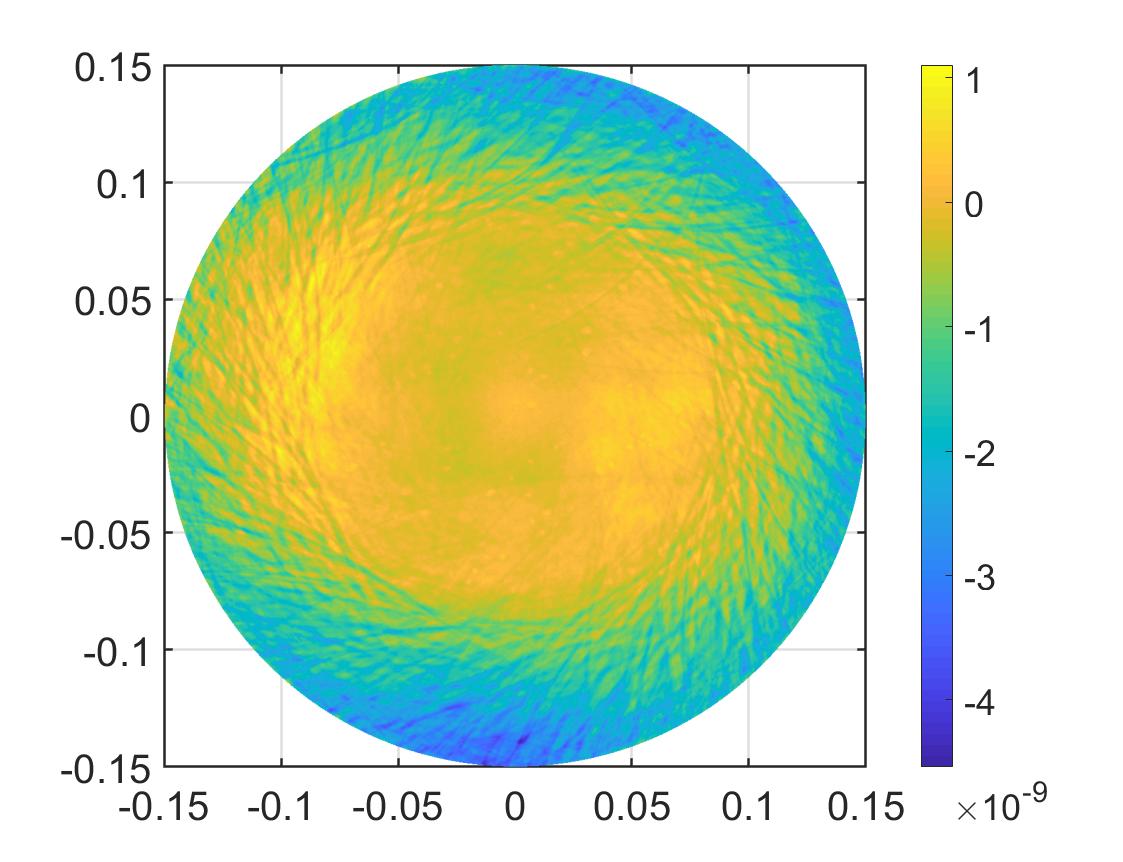}	
	\includegraphics[width=0.325\textwidth]{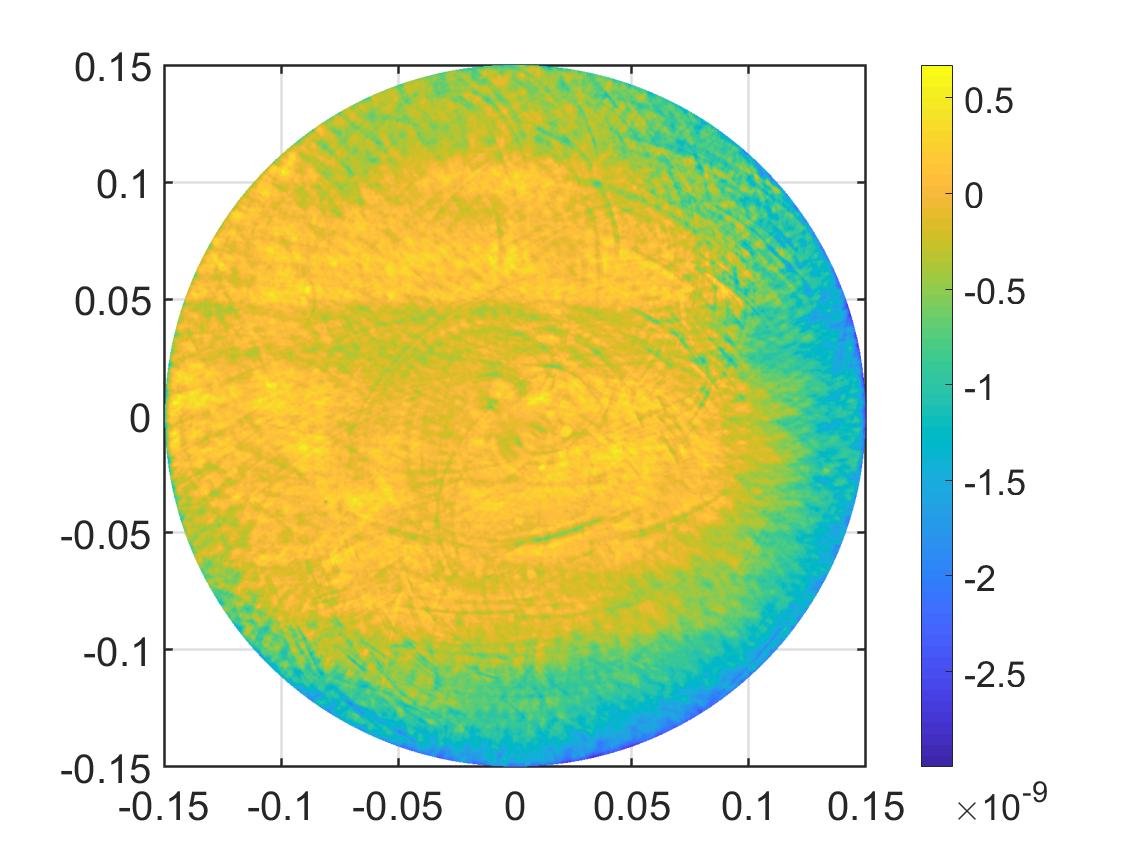}
 \caption{Three maps of roughness of the LIGO mirrors surface  which were utilized in the simulations. X-axis and Y-axis are presented in meters.}\label{maps}
\end{figure*}

Stability of the modes in the selected cavity are very important. Real cavity mirrors cannot be perfectly smooth but have coordinate-dependent surface profile $\beta(r, \varphi)$ with roughness of the order of several nanometers (for the LIGO mirrors). Several examples of the mirrors are shown in Fig.~\ref{maps}. To verify stability of the solution in such a cavity, we select mirrors defined in the set 1 of Table \ref{EV_problem_asph}, add one of these roughness maps to one of the mirrors, and evaluate the optical losses of the main mode of the cavity. The simulation shows that the associated increase of the loss is insignificant. Hence, the solution is stable with respect of the nonzero roughness. 

The roughness map does not have cylindrical symmetry of the cavity. At first site the problem cannot be reduced to two dimensions (\ref{EV}) and should be evaluated in all three dimensions. However, modeling cavity in three dimensions is impractical and extremely time consuming. We apply 2D method of successive approximation to estimate which part of fundamental mode will scatter on the roughness map (in other two dimensions $\rho, \phi$) in other mode families (with non-zero azimuthal index $\ell \neq 0$). In other words, certain 3D pattern on a mirror might lead to coupling between two degenerate modes belonging to two different mode families. We start from 
%
\begin{align}
\label{tildePsi}
	\widetilde{\lambda}^2_{n, \ell}\widetilde{\Psi}_{n, \ell} (\vec{x}_1) = \int g(\vec{x}_1, \vec{x}_2)e^{2i\beta}\widetilde{\Psi}_{n, \ell} (\vec{x}_2)d\vec{x}_2,
\end{align}

Following \citep{18PoMaYaVy} we utilize a decomposition
\begin{align}
	e^{-2i\beta} &= \delta_1 + \delta_2 + \delta_3 +\dots, \\	
	\tilde\lambda_{00}^2 &= \lambda_{00 }^2 + \left(\tilde\lambda_{00 }^2\right)^{(1)} + \left(\tilde\lambda_{00}^2\right)^{(2)}+\dots,\\
	\left(\tilde\lambda_{00 }^2\right)^{(1)}  &= \lambda_{00}^2 V^{(1)}_{00,00 }, \\
	V_{00,00}^{(1)}&\equiv \int \psi^*_{00 }(x_1) \,\delta_1(\vec x_1)\, \psi_{00 }(x_1) \,x_1\,d x_1\, d\phi_1 ,\\
	\left(\tilde\lambda_{00 }^2\right)^{(2)}  &= \lambda_{00}^2\Delta V,\\
	 \Delta V \equiv & \left( V^{(2)}_{00,00 } -  2\sum_{
	 k, m \in \mathbf{Z}_+}  \frac{\lambda_{km}^2}{\lambda_{00}^2 - \lambda_{km}^2}\cdot\left| V_{km,00}^{(1)} \right|^2\right),
	 \label{deltaV}
\end{align}
to solve Eq.(\ref{tildePsi}).

We used the method for two cavities with nonidentical spherical (LIGO parameters) and non-spherical mirrors (parameter set 1). The mode was propagated by 300~m. A few real roughness maps $\beta(r, \varphi)$ were utilized. The loss increased from 0.5 to 2 ppm for both types of mirrors. No significant increase of the loss was recorded and, hence, the solution is stable with respect to the surface perturbations.
 
The numerical procedure has the following steps:

Step 1. Modes of a symmetric cavity were shifted by a distance of 300~m, phase profiles of the modes were obtained; the phase profiles were then used as mirrors of an asymmetric cavity. 

Step 2. All the eigenmodes (field distributions and eigenvalues with the diffraction loss not exceeding $10^4$ ppm per a round trip) were found.

Step 3. The roughness of spherical profile $\beta(r, \varphi)$ of an aLIGO mirror (measured in meters after tilt and curvature subtraction using a standard mathematical approach) was added to the mirrors.

Step 4. The grids were matched. Due to the fact that the map of roughness was formed by measuring the surface over the same distance, there is a small difficulty with the calculation of integrals, since the own modes of cavities have their own grid of coordinates, tied to the roots of the equation for each azimuthal index $\ell$ (\ref{eq_root}). Therefore it is necessary to bring each grid of coordinates together with own modes to a grid of coordinates of the roughness maps. For this purpose, the standard Matlab function ``interp1'' with the method ``spline'' was used.

Step 5. The calculation of the ``overlap'' integral for each pair of modes $\Psi_{00}$ and $\Psi_{km}$ and further calculation of diffraction losses through the perturbed eigenvalue of the main mode $\tilde{\mathcal{L}_{00}} = 1 - \big | \tilde{\lambda_{00}} \big |^2$ took place.

For a cavity with non-identical spherical mirrors after mode propagation at the distance of 300~m for the three real maps of the roughness  \cite{LIGO_mirrors} (maps ETM07, ETM08, ETM09 available in open access) presented Fig.~\ref{maps}  we obtained the following additional contributions to the diffraction loss $\Delta \mathcal{L}_{00} = \tilde{\mathcal{L}_{00}} - \mathcal{L}_{00}$:
\begin{align}
	&\Delta \mathcal{L}_{00}^{sph, 1} = 0.48 \text{ ppm}, \\
	&\Delta \mathcal{L}_{00}^{sph, 2} = 1.49 \text{ ppm}, \\
	&\Delta \mathcal{L}_{00}^{sph, 3} = 2.12 \text{ ppm}
\end{align}

For a cavity with non-identical non-spherical mirrors (parameter set 1) after a 300~m mode propagation, the following values were obtained (Fig.~\ref{maps}):

\begin{align}
	&\Delta \mathcal{L}_{00}^{set1, 1} = 0.48 \text{ ppm}, \\ 	
	&\Delta \mathcal{L}_{00}^{set1, 2} = 1.28 \text{ ppm}, \\ 	
	&\Delta \mathcal{L}_{00}^{set1, 3} = 1.98 \text{ ppm}
\end{align}

The additional loss does not depend on changes in the shape of the mirrors due to the similarity of the profile of the main modes of the Fabry-Perot cavities with spherical and non-spherical mirrors.

\section{Conclusions}

In this article we introduce a multipurpose technique of numerical evaluation of diffraction loss in optical cavities with non-identical mirrors. The method is based on the imaginative free propagation of the beams beyond the mirrors of a cavity having identical mirrors. The phase distributions of the beams define the shapes of the mirrors. The method allows us to simplify the calculation since the cavity with identical mirrors can be simulated rather fast. We utilized a cavity with spherical mirrors to calibrate the technique and used it to design a cavity with non-identical non-spherical mirrors. The cavity allows to reduce the quality factors of all but one optical mode families and suppress the parametric instability involving the modes.

\acknowledgments

M.P. and S.V. acknowledges support from  Russian Foundation of Basic Research (Grant No. 19-29-11003) and from the TAPIR GIFT MSU Support of California Institute of Technology. The reported here research performed by A.M was carried out at the Jet Propulsion Laboratory, California Institute of Technology, under a contract with the National Aeronautics and Space Administration (80NM0018D0004). The LIGO Observatories were constructed by the California Institute of Technology and Massachusetts Institute of Technology with funding from the National Science Foundation under cooperative agreement PHY-9210038. The LIGO Laboratory operates under cooperative agreement PHY-1764464.  This paper carries LIGO Document Number LIGO-P2000138.

\section*{Appendix A: X-axis normalization}

Dimensionless coordinate $x$ depends on length $L$ of the cavity (\ref{def1}). This is an inconvenient choice for the mode propagation method as $L$ is changing. For example let us consider an eigenmode of the cavity 1-2 shown in Fig.~\ref{cav}. The filed distribution is given by function $\Psi_2 \left(\frac{r}{b}\right)$ with $x=r/b$ and $b=\sqrt {L/k}$ \eqref{def1}. 

At the distance $L' = L + \Delta L$ (mirror 6) the x-axis should be renormalized for any calculations with shifted mode:
\begin{align}
	\nonumber
	x = \frac{r}{b}, \quad b = \sqrt{\frac{L}{k}} \quad & \rightarrow \quad x' = \frac{r}{b'}, \quad b' = \sqrt{\frac{L'}{k}}
\end{align}
The result is
\begin{subequations}
 \label{G'}
\begin{align}
	\tilde{\Psi}_6(x') &= \textbf{P}(L') \textbf{R}\Psi_2(x'), \\
	\textbf{P}(L') &= \left(\textbf{H}^{(+)}\right)^{-1} \tilde{\textbf G}(L')\mathbf{H}^{(+)}, \\
	\tilde{\textbf G}_{\alpha\beta} (L') &= \exp\left(-\frac{i}{2}\cdot \frac{\xi_\alpha^2}{a^2}\frac{L'}{L}\right)\delta_{\alpha\beta}
\end{align}
\end{subequations}
The matrix $\tilde{\mathbf G}$ depends on $L'/L$ while matrices $\mathbf{H}^{(\pm)}$ do not (see \eqref{Hpm}).

It is more convenient to use normalization related to the initial conditions instead. The distribution functions are recalculated using normalization $\int |\Psi(x)|^2 \, x\, dx=1$:
\begin{align}
  \label{recalc}
  \tilde{\Psi}_6(x')= \frac{b}{b'}\cdot \tilde{\Psi}_6(x),\quad 
    \tilde{\Psi}_2(x')= \frac{b}{b'}\cdot \tilde{\Psi}_2(x)
\end{align}
We substitute  $\Psi(x),\ \tilde{\Psi}_6(x)$ into \eqref{G'} instead of $\Psi(x'),\ \tilde{\Psi}_6(x')$ (factors $b/b'$ are reduced). For example, to calculate the radius of curvature (in the case of spherical mirrors), one can use two equivalent  formulas 
\begin{align}
	R_{c6} = \frac{x'^2 L'}{2 \text{arg\,}\big(\tilde{\Psi}_6(x')\big)},\quad \text{or}\quad
	R_{c6} = \frac{x^2 L}{2 \text{arg\,}\big(\tilde{\Psi}_6(x)\big)}
\end{align}

Let us consider a mode propagation  from mirror 2 to mirror 3 at the distance $\Delta L$. Formally, x-axis should be renormalized to the new distance $L''=\Delta L$. 
\begin{align}
	x'' = \frac{r}{b''}, \quad b'' = \sqrt{\frac{L''}{k}}.
\end{align}
The distribution $\Psi_3$ can be calculated by \eqref{G'} with matrix $\tilde{\mathbf G}''$ corrected correspondingly ($L'' \to L'$).

\begin{figure}[h]
	\includegraphics[width=0.49\textwidth]{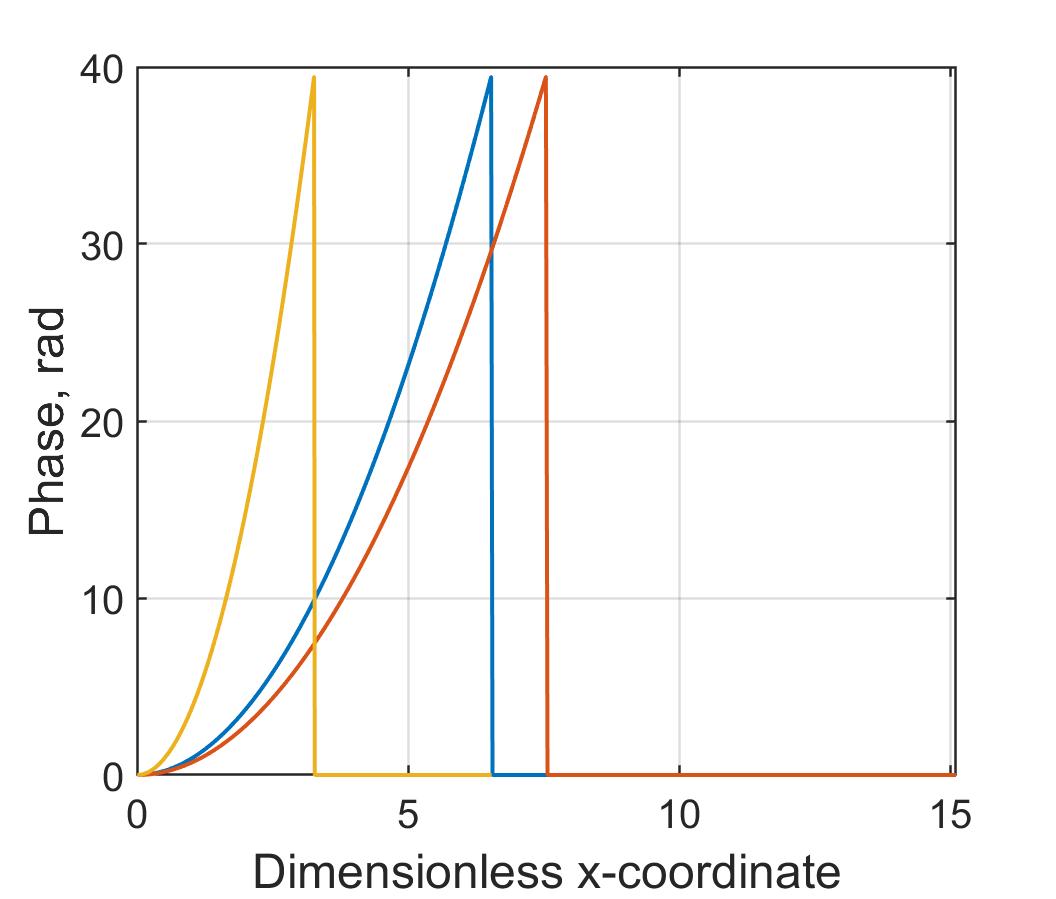}
 \caption{Phase distribution of $\Psi_6(x')$ after mode propagation from mirror 2 to 6 (red line) and $\Psi_6(x'')$ from mirror 1 to mirror 6 (yellow line) plotted on different $x'$ and $x''$-axes correspondingly. We have selected here $\Delta L = 1000$~m. Blue line corresponds to phase distribution $\Psi_6(x)$ in x-axis of length $L$.}\label{Renorm}
\end{figure}

The plots on Fig.~\ref{Renorm} illustrate the phase of distribution $\tilde{\Psi}_6$ with different normalization utilized.

Despite the difference in the dimensionless coordinates, the point with the index $M \simeq N/S$ of the each phase distribution (among them $\text{arg\,}(\tilde{\Psi}_3), \text{arg\,}(\tilde{\Psi}_4)$) on the mirror corresponds to the edge of the mirror $a_m = a/S$. Please note that x-axis is defined though the roots of the characteristic equation (\ref{x-axis}). Points of different x-axis $x(M) = \frac{r(M)}{b}$, $x'(M) = \frac{r(M)}{b'}$, $x''(M) = \frac{r(M)}{b''}$, ... also correspond to the edge of the mirror and to the root $\xi(M)$. 

To conclude, there is no need to renormalize dimensionless coordinates $x$ to new distance $L'$ or $L''$ if we correct the propagation matrix $\tilde {\mathbf G'}$ as well as normalize the values of the distributions \eqref{recalc}.

\section*{Appendix B: Fitting by exponential function}

The results of the simulations of the mode propagation on 300~m and further re-optimization of obtained phase distribution are presented below. The phase distribution was fitted by the Eq. (\ref{ynew}) using standard MatLab tool ``polyfit''. Values of coefficients $c_i$ and $\tilde{h}_0$ after re-optimization are shown in table \ref{coef}. 

Note that parameters sets 1, 2, 3 were used with normalization on 3700~m instead of 4000~m. 

\begin{table}[h!]
\centering
\begin{tabular}{|c|c|c|c|}
\hline 
• & set 1 & set 2 & set 3 \\ 
\hline
$\tilde{h}_0$ & 44 & 31.8 & 52.1 \\ 
\hline  
$c_1$ & -1.36e-5 & -4.78e-6& -8.31e-6 \\ 
\hline 
$c_2$ & 1.62e-4 & 1.48e-4 & 1.24e-4 \\ 
\hline 
$c_3$ & -8.07e-4 & -1.05e-3 & -8.01e-4 \\ 
\hline 
$c_4$ & 2.15e-3 & 4.05e-3 & 2.27e-3 \\ 
\hline 
$c_5$ & 1.9e-2 & 2.48e-2 & 1.58e-2 \\ 
\hline 
$c_6$ & 3.13e-4 & 1.16e-3 & 2.46e-4 \\ 
\hline 
\end{tabular}
\caption{Coefficients of the approximation applied after the re-optimization. Each column corresponds to the second mirror in the new non-symmetrical cavity, in which the first one is described by the set of parameters from the first cell.} \label{coef}
\end{table}

\bibliography{mike}

\begin{thebibliography}{28}
\expandafter\ifx\csname natexlab\endcsname\relax\def\natexlab#1{#1}\fi
\expandafter\ifx\csname bibnamefont\endcsname\relax
  \def\bibnamefont#1{#1}\fi
\expandafter\ifx\csname bibfnamefont\endcsname\relax
  \def\bibfnamefont#1{#1}\fi
\expandafter\ifx\csname citenamefont\endcsname\relax
  \def\citenamefont#1{#1}\fi
\expandafter\ifx\csname url\endcsname\relax
  \def\url#1{\texttt{#1}}\fi
\expandafter\ifx\csname urlprefix\endcsname\relax\def\urlprefix{URL }\fi
\providecommand{\bibinfo}[2]{#2}
\providecommand{\eprint}[2][]{\url{#2}}

\bibitem[{\citenamefont{Ferdous et~al.}(2014)\citenamefont{Ferdous, Demchenko,
  Vyatchanin, Matsko, and Maleki}}]{14FeDeVy}
\bibinfo{author}{\bibfnamefont{F.}~\bibnamefont{Ferdous}},
  \bibinfo{author}{\bibfnamefont{A.}~\bibnamefont{Demchenko}},
  \bibinfo{author}{\bibfnamefont{S.}~\bibnamefont{Vyatchanin}},
  \bibinfo{author}{\bibfnamefont{A.}~\bibnamefont{Matsko}}, \bibnamefont{and}
  \bibinfo{author}{\bibfnamefont{L.}~\bibnamefont{Maleki}},
  \bibinfo{journal}{Physical Review A} \textbf{\bibinfo{volume}{90}},
  \bibinfo{pages}{033826} (\bibinfo{year}{2014}).

\bibitem[{\citenamefont{Matsko et~al.}(2016)\citenamefont{Matsko, Poplavskiy,
  Yamamoto, and Vyatchanin}}]{16MaPoYa}
\bibinfo{author}{\bibfnamefont{A.}~\bibnamefont{Matsko}},
  \bibinfo{author}{\bibfnamefont{M.}~\bibnamefont{Poplavskiy}},
  \bibinfo{author}{\bibfnamefont{H.}~\bibnamefont{Yamamoto}}, \bibnamefont{and}
  \bibinfo{author}{\bibfnamefont{S.}~\bibnamefont{Vyatchanin}},
  \bibinfo{journal}{Physical Review D} \textbf{\bibinfo{volume}{93}},
  \bibinfo{pages}{083010} (\bibinfo{year}{2016}).

\bibitem[{\citenamefont{Abbott and et~al}(2016)}]{GW150914}
\bibinfo{author}{\bibfnamefont{B.}~\bibnamefont{Abbott}} \bibnamefont{and}
  \bibinfo{author}{\bibnamefont{et~al}}, \bibinfo{journal}{Physical review
  letters} \textbf{\bibinfo{volume}{116 (6)}}, \bibinfo{pages}{061102}
  (\bibinfo{year}{2016}).

\bibitem[{\citenamefont{Abbott and et~al}(2017{\natexlab{a}})}]{GW170104}
\bibinfo{author}{\bibfnamefont{B.}~\bibnamefont{Abbott}} \bibnamefont{and}
  \bibinfo{author}{\bibnamefont{et~al}}, \bibinfo{journal}{Physical review
  letters} \textbf{\bibinfo{volume}{118 (22)}}, \bibinfo{pages}{221101}
  (\bibinfo{year}{2017}{\natexlab{a}}).

\bibitem[{\citenamefont{Abbott and et~al}(2017{\natexlab{b}})}]{GW170814}
\bibinfo{author}{\bibfnamefont{B.}~\bibnamefont{Abbott}} \bibnamefont{and}
  \bibinfo{author}{\bibnamefont{et~al}}, \bibinfo{journal}{Physical review
  letters} \textbf{\bibinfo{volume}{119 (14)}}, \bibinfo{pages}{141101}
  (\bibinfo{year}{2017}{\natexlab{b}}).

\bibitem[{\citenamefont{Abbott and et~al}(2018{\natexlab{a}})}]{GW170817}
\bibinfo{author}{\bibfnamefont{B.}~\bibnamefont{Abbott}} \bibnamefont{and}
  \bibinfo{author}{\bibnamefont{et~al}}, \bibinfo{journal}{Physical review
  letters} \textbf{\bibinfo{volume}{121 (16)}}, \bibinfo{pages}{161101}
  (\bibinfo{year}{2018}{\natexlab{a}}).

\bibitem[{\citenamefont{Acernese and et~al}(2015)}]{virgo}
\bibinfo{author}{\bibfnamefont{F.}~\bibnamefont{Acernese}} \bibnamefont{and}
  \bibinfo{author}{\bibnamefont{et~al}}, \bibinfo{journal}{Classical and
  Quantum Gravity} \textbf{\bibinfo{volume}{32}}, \bibinfo{pages}{24001}
  (\bibinfo{year}{2015}).

\bibitem[{\citenamefont{Somiya and et~al}(2012)}]{kagra2}
\bibinfo{author}{\bibfnamefont{K.}~\bibnamefont{Somiya}} \bibnamefont{and}
  \bibinfo{author}{\bibnamefont{et~al}}, \bibinfo{journal}{Classical and
  Quantum Gravity} \textbf{\bibinfo{volume}{29}}, \bibinfo{pages}{124007}
  (\bibinfo{year}{2012}).

\bibitem[{\citenamefont{Aso and et~al}(2013)}]{kagra1}
\bibinfo{author}{\bibfnamefont{Y.}~\bibnamefont{Aso}} \bibnamefont{and}
  \bibinfo{author}{\bibnamefont{et~al}}, \bibinfo{journal}{Phys. Rev. D}
  \textbf{\bibinfo{volume}{88}}, \bibinfo{pages}{043007}
  (\bibinfo{year}{2013}).

\bibitem[{\citenamefont{Abbott and et~al}(2014)}]{pfaLIGO}
\bibinfo{author}{\bibfnamefont{B.}~\bibnamefont{Abbott}} \bibnamefont{and}
  \bibinfo{author}{\bibnamefont{et~al}}, \bibinfo{journal}{arXiv.1411.4547}
  (\bibinfo{year}{2014}).

\bibitem[{\citenamefont{Abbott and et~al}(2017{\natexlab{c}})}]{17LIGO_1}
\bibinfo{author}{\bibfnamefont{B.}~\bibnamefont{Abbott}} \bibnamefont{and}
  \bibinfo{author}{\bibnamefont{et~al}}, \bibinfo{journal}{Classical and
  Quantum Gravity} \textbf{\bibinfo{volume}{34 (4)}}, \bibinfo{pages}{044001}
  (\bibinfo{year}{2017}{\natexlab{c}}).

\bibitem[{\citenamefont{Abbott and et~al}(2018{\natexlab{b}})}]{18LIGO_1}
\bibinfo{author}{\bibfnamefont{B.}~\bibnamefont{Abbott}} \bibnamefont{and}
  \bibinfo{author}{\bibnamefont{et~al}}, \bibinfo{journal}{Living Reviews in
  Relativity} \textbf{\bibinfo{volume}{21}}
  (\bibinfo{year}{2018}{\natexlab{b}}).

\bibitem[{\citenamefont{Braginsky et~al.}(2001)\citenamefont{Braginsky,
  Strigin, and Vyatchanin}}]{01BrStVy}
\bibinfo{author}{\bibfnamefont{V.}~\bibnamefont{Braginsky}},
  \bibinfo{author}{\bibfnamefont{S.}~\bibnamefont{Strigin}}, \bibnamefont{and}
  \bibinfo{author}{\bibfnamefont{S.}~\bibnamefont{Vyatchanin}},
  \bibinfo{journal}{Physics letters A} \textbf{\bibinfo{volume}{287}},
  \bibinfo{pages}{331} (\bibinfo{year}{2001}).

\bibitem[{\citenamefont{Braginsky et~al.}(2002)\citenamefont{Braginsky,
  Strigin, and Vyatchanin}}]{02BrStVy}
\bibinfo{author}{\bibfnamefont{V.}~\bibnamefont{Braginsky}},
  \bibinfo{author}{\bibfnamefont{S.}~\bibnamefont{Strigin}}, \bibnamefont{and}
  \bibinfo{author}{\bibfnamefont{S.}~\bibnamefont{Vyatchanin}},
  \bibinfo{journal}{Physics letters A} \textbf{\bibinfo{volume}{305}},
  \bibinfo{pages}{111} (\bibinfo{year}{2002}).

\bibitem[{\citenamefont{Chen et~al.}(2015)\citenamefont{Chen, Zhao, Danilishin,
  and et~al}}]{15ChZhDa}
\bibinfo{author}{\bibfnamefont{X.}~\bibnamefont{Chen}},
  \bibinfo{author}{\bibfnamefont{C.}~\bibnamefont{Zhao}},
  \bibinfo{author}{\bibfnamefont{S.}~\bibnamefont{Danilishin}},
  \bibnamefont{and} \bibinfo{author}{\bibnamefont{et~al}},
  \bibinfo{journal}{Physical Review A} \textbf{\bibinfo{volume}{91}},
  \bibinfo{pages}{033832} (\bibinfo{year}{2015}).

\bibitem[{\citenamefont{Kippenberg et~al.}(2005)\citenamefont{Kippenberg,
  Rokhsari, Carmon, Scherer, and Vahala}}]{05KipVah}
\bibinfo{author}{\bibfnamefont{T.}~\bibnamefont{Kippenberg}},
  \bibinfo{author}{\bibfnamefont{H.}~\bibnamefont{Rokhsari}},
  \bibinfo{author}{\bibfnamefont{T.}~\bibnamefont{Carmon}},
  \bibinfo{author}{\bibfnamefont{A.}~\bibnamefont{Scherer}}, \bibnamefont{and}
  \bibinfo{author}{\bibfnamefont{K.}~\bibnamefont{Vahala}},
  \bibinfo{journal}{Physical Review Letters} \textbf{\bibinfo{volume}{95}},
  \bibinfo{pages}{033901} (\bibinfo{year}{2005}).

\bibitem[{\citenamefont{M.Evans et~al.}(2010)\citenamefont{M.Evans, L.Barsotti,
  and P.Fritschel}}]{EvansPLA2010}
\bibinfo{author}{\bibnamefont{M.Evans}},
  \bibinfo{author}{\bibnamefont{L.Barsotti}}, \bibnamefont{and}
  \bibinfo{author}{\bibnamefont{P.Fritschel}}, \bibinfo{journal}{Physics
  Letters A} \textbf{\bibinfo{volume}{374}}, \bibinfo{pages}{665}
  (\bibinfo{year}{2010}).

\bibitem[{\citenamefont{Evans and et~al}(2015)}]{15EvGrFr}
\bibinfo{author}{\bibfnamefont{M.}~\bibnamefont{Evans}} \bibnamefont{and}
  \bibinfo{author}{\bibnamefont{et~al}}, \bibinfo{journal}{Physical Review
  Letters} \textbf{\bibinfo{volume}{114}}, \bibinfo{pages}{161102}
  (\bibinfo{year}{2015}).

\bibitem[{\citenamefont{Degallaix et~al.}(2007)\citenamefont{Degallaix, Zhao,
  Ju, and Blair}}]{07DeZhJu}
\bibinfo{author}{\bibfnamefont{J.}~\bibnamefont{Degallaix}},
  \bibinfo{author}{\bibfnamefont{C.}~\bibnamefont{Zhao}},
  \bibinfo{author}{\bibfnamefont{L.}~\bibnamefont{Ju}}, \bibnamefont{and}
  \bibinfo{author}{\bibfnamefont{D.}~\bibnamefont{Blair}},
  \bibinfo{journal}{JOSA B} \textbf{\bibinfo{volume}{24(6)}},
  \bibinfo{pages}{13361343} (\bibinfo{year}{2007}).

\bibitem[{\citenamefont{Fan et~al.}(2010)\citenamefont{Fan, Merrill, Zhao, Ju,
  Blair, Slagmolen, Hosken, Brooks, Veitch, and Munch}}]{10FaMeZh}
\bibinfo{author}{\bibfnamefont{Y.}~\bibnamefont{Fan}},
  \bibinfo{author}{\bibfnamefont{L.}~\bibnamefont{Merrill}},
  \bibinfo{author}{\bibfnamefont{C.}~\bibnamefont{Zhao}},
  \bibinfo{author}{\bibfnamefont{L.}~\bibnamefont{Ju}},
  \bibinfo{author}{\bibfnamefont{D.}~\bibnamefont{Blair}},
  \bibinfo{author}{\bibfnamefont{B.}~\bibnamefont{Slagmolen}},
  \bibinfo{author}{\bibfnamefont{D.}~\bibnamefont{Hosken}},
  \bibinfo{author}{\bibfnamefont{A.}~\bibnamefont{Brooks}},
  \bibinfo{author}{\bibfnamefont{P.}~\bibnamefont{Veitch}}, \bibnamefont{and}
  \bibinfo{author}{\bibfnamefont{J.}~\bibnamefont{Munch}},
  \bibinfo{journal}{Classical and Quantum Gravity}
  \textbf{\bibinfo{volume}{27}}, \bibinfo{pages}{084028}
  (\bibinfo{year}{2010}).

\bibitem[{\citenamefont{Miller et~al.}(2011)\citenamefont{Miller, Evans,
  Barsotti, Fritschel, MacInnis, Mittleman, Shapiro, Soto, and
  Torrie}}]{11MiEfBa}
\bibinfo{author}{\bibfnamefont{J.}~\bibnamefont{Miller}},
  \bibinfo{author}{\bibfnamefont{M.}~\bibnamefont{Evans}},
  \bibinfo{author}{\bibfnamefont{L.}~\bibnamefont{Barsotti}},
  \bibinfo{author}{\bibfnamefont{P.}~\bibnamefont{Fritschel}},
  \bibinfo{author}{\bibfnamefont{M.}~\bibnamefont{MacInnis}},
  \bibinfo{author}{\bibfnamefont{R.}~\bibnamefont{Mittleman}},
  \bibinfo{author}{\bibfnamefont{B.}~\bibnamefont{Shapiro}},
  \bibinfo{author}{\bibfnamefont{J.}~\bibnamefont{Soto}}, \bibnamefont{and}
  \bibinfo{author}{\bibfnamefont{C.}~\bibnamefont{Torrie}},
  \bibinfo{journal}{Physics Letters A} \textbf{\bibinfo{volume}{375}},
  \bibinfo{pages}{788} (\bibinfo{year}{2011}).

\bibitem[{\citenamefont{Biscans et~al.}(2019)\citenamefont{Biscans, Gras,
  Blair, Driggers, Evans, Fritschel, Hardwick, and Mansel}}]{LIGO_damper}
\bibinfo{author}{\bibfnamefont{S.}~\bibnamefont{Biscans}},
  \bibinfo{author}{\bibfnamefont{S.}~\bibnamefont{Gras}},
  \bibinfo{author}{\bibfnamefont{C.~D.} \bibnamefont{Blair}},
  \bibinfo{author}{\bibfnamefont{J.}~\bibnamefont{Driggers}},
  \bibinfo{author}{\bibfnamefont{M.}~\bibnamefont{Evans}},
  \bibinfo{author}{\bibfnamefont{P.}~\bibnamefont{Fritschel}},
  \bibinfo{author}{\bibfnamefont{T.}~\bibnamefont{Hardwick}}, \bibnamefont{and}
  \bibinfo{author}{\bibfnamefont{G.}~\bibnamefont{Mansel}},
  \bibinfo{journal}{Physical Review D} \textbf{\bibinfo{volume}{100}},
  \bibinfo{pages}{122003} (\bibinfo{year}{2019}).

\bibitem[{\citenamefont{Buikema}(2019)}]{PA_aLIGO}
\bibinfo{author}{\bibfnamefont{A.}~\bibnamefont{Buikema}},
  \bibinfo{journal}{Amaldi 13, LIGO-G1900455}  (\bibinfo{year}{2019}),
  \urlprefix\url{https://dcc.ligo.org/LIGO-G1900455/public}.

\bibitem[{\citenamefont{Allocca et~al.}(2019)\citenamefont{Allocca, Chiummo,
  Ruggi, and Yamamoto}}]{PA_AdVirgo}
\bibinfo{author}{\bibfnamefont{A.}~\bibnamefont{Allocca}},
  \bibinfo{author}{\bibfnamefont{A.}~\bibnamefont{Chiummo}},
  \bibinfo{author}{\bibfnamefont{P.}~\bibnamefont{Ruggi}}, \bibnamefont{and}
  \bibinfo{author}{\bibfnamefont{H.}~\bibnamefont{Yamamoto}},
  \bibinfo{journal}{VIR-1047A-19}  (\bibinfo{year}{2019}),
  \urlprefix\url{https://tds.virgo-gw.eu/ql/?c=14881}.

\bibitem[{\citenamefont{Yamamoto}(2020)}]{PA_HY}
\bibinfo{author}{\bibfnamefont{H.}~\bibnamefont{Yamamoto}},
  \bibinfo{journal}{LIGO-G2000282}  (\bibinfo{year}{2020}),
  \urlprefix\url{https://dcc.ligo.org/LIGO-G2000282/public}.

\bibitem[{\citenamefont{Vinet and Hello}(1993)}]{93vinet}
\bibinfo{author}{\bibfnamefont{J.}~\bibnamefont{Vinet}} \bibnamefont{and}
  \bibinfo{author}{\bibfnamefont{P.}~\bibnamefont{Hello}},
  \bibinfo{journal}{Journal of Modern Optics} \textbf{\bibinfo{volume}{40}},
  \bibinfo{pages}{1981} (\bibinfo{year}{1993}).

\bibitem[{\citenamefont{Poplavskiy et~al.}(2018)\citenamefont{Poplavskiy,
  Matsko, Yamamoto, and Vyatchanin}}]{18PoMaYaVy}
\bibinfo{author}{\bibfnamefont{M.}~\bibnamefont{Poplavskiy}},
  \bibinfo{author}{\bibfnamefont{A.}~\bibnamefont{Matsko}},
  \bibinfo{author}{\bibfnamefont{H.}~\bibnamefont{Yamamoto}}, \bibnamefont{and}
  \bibinfo{author}{\bibfnamefont{S.}~\bibnamefont{Vyatchanin}},
  \bibinfo{journal}{Journal of Optics} \textbf{\bibinfo{volume}{20}},
  \bibinfo{pages}{075609} (\bibinfo{year}{2018}).

\bibitem[{LIG()}]{LIGO_mirrors}
\urlprefix\url{https://galaxy.ligo.caltech.edu/optics/}.

\end{thebibliography}

\end{document}